\documentclass[12pt]{iopart}
\pdfoutput=1
\usepackage{iopams}
\usepackage{ulem}
 
\expandafter\let\csname equation*\endcsname\relax
\expandafter\let\csname endequation*\endcsname\relax
\usepackage{amsmath,amssymb,mathtools}

\usepackage[colorlinks,linkcolor=blue,urlcolor=blue,citecolor=blue]{hyperref}

\def\be{\begin{equation}}
\def\ee{\end{equation}}
\def\bea{\begin{eqnarray}}
\def\eea{\end{eqnarray}}

\usepackage{xcolor}
\definecolor{dgreen}{rgb}{0,0.7,0}

\begin{document}

\title{The inspection paradox in stochastic resetting}

\author{Arnab Pal$^{1,2}$, Sarah Kostinski$^{1}$, and Shlomi Reuveni$^{1}$\footnote{arnabp@iitk.ac.in; skostinski@tauex.tau.ac.il; shlomire@tauex.tau.ac.il}}
\address{\textit{$^{1}$School of Chemistry, The Center for Physics and Chemistry of Living Systems, The Raymond and Beverly Sackler Center for Computational Molecular and Materials Science, \& The Mark Ratner Institute for Single Molecule Chemistry, Tel Aviv University, Tel Aviv 6997801, Israel}}
\address{\textit{$^{2}$Department of Physics, Indian Institute of Technology, Kanpur, Kanpur 208016, India}}
\date{\today}

\begin{abstract} The remaining travel time of a plane shortens with every minute that passes from its departure, and a flame diminishes a candle with every second it burns. Such everyday occurrences bias us to think that processes which have already begun will end before those which have just started. Yet, the inspection paradox teaches us that the converse can also happen when randomness is at play. The paradox comes from probability theory, where it is often illustrated by measuring how long passengers wait upon arriving at a bus stop at a random time. 
Interestingly, such passengers may on average wait longer than the mean time between bus arrivals -- a counter-intuitive result, since one expects to wait less when coming some time after the previous bus departed. In this \textit{viewpoint}, we review the inspection paradox and its origins. The insight gained is then used to explain why, in some situations, stochastic resetting expedites the completion of random processes. Importantly, this is done with elementary mathematical tools which help develop a probabilistic intuition for stochastic resetting and how it works. This viewpoint can thus be used as an accessible  introduction to the subject.
\end{abstract}

\section{Introduction}
\label{introduction}
The field of stochastic `resetting,' or `restart,' has recently emerged from the investigation of various nonequilibrium systems in physics \cite{Restart1,Restart2,Restart3,Levy,Kirone,Pal-potential,transport1,transport2,Pal-time-dep,powerlaw-diff,Durang,path,ReuveniPRL,PalReuveniPRL,PalReuveniBranching,anamolous-1,anamolous-3,RAP,scaled,CTRW1,CTRW2,CTRW3,CTRW4,Levy-flight,telegraphic,comb,Ising,KPZ,SEP,SEP-2,localex,TASEP,localt,accm,occ,cost,cost2,Chechkin,Belan,Belan-2,sth-0,sth-1,sth-2,sth-3,evs-1,evs-2,evs-3,non-inst-det-1,non-inst-det-2,non-inst-det-3,non-inst-det-4,non-inst-det-5,non-inst-det-6,non-inst-st-1,non-inst-st-2,non-inst-st-3,nonlinear,random-amplitude,active-L,gated-search-resetting,gated-search-resetting-2,underdamped,ldf-1,ldf-2,fbm-erg,diff-erg,cat,RW-m,evapor,frac-der,ZRP,DP}, chemistry \cite{Restart-Biophysics1,Restart-Biophysics2,Restart-Biophysics3,Restart-Biophysics4}, biology \cite{Restart-Bio-1,Restart-Bio-2,Restart-Bio-3,Restart-Bio-4,RT,RTP,RTP-2d}, ecology \cite{HRS,Montanari,bressloff}, queuing theory \cite{q1,q2,q3,q4}, computer science \cite{Luby,algorithm,algorithm-2,c1,c2,c3} and economics \cite{eco-1,eco-2,eco-3}. See \cite{review} for an overview of the subject. One might think that repeatedly restarting the dynamics of a stochastic process only hinders its completion, but surprisingly this is not always the case. Indeed, a hallmark property of restart is its ability to expedite the completion of certain processes.  For example, restart can decrease the mean first-passage time of a diffusing particle to reach a boundary or target~\cite{Restart1,ReuveniPRL,PalReuveniPRL,expt-1,expt-2,expt-3}, the mean run time of stochastic computer algorithms~\cite{c1,c2,c3}, the mean turnover time of enzymatic reactions~\cite{Restart-Biophysics1,Restart-Biophysics2}, and the mean search time of foraging animals~\cite{HRS}.  Restart is therefore a useful means to regulate completion of stochastic processes and dynamical systems (see \cite{Restart1,PalReuveniPRL,Chechkin} and additional references in \cite{review}). 

However, it is not immediately clear under which conditions restart will actually expedite completion of a stochastic process, as there are also cases where it causes significant slowdown. To this end, a series of universal relations were derived, determining such criteria~\cite{ReuveniPRL,PalReuveniPRL,PalReuveniBranching,Chechkin,HRS,sharp-1,sharp-2,sharp-3}.  The derivations of these `restart-criteria' typically rely on the renewal structure of resetting and other non-elementary mathematical tools, which are less accessible to students and readers outside the field.  The purpose of this \textit{viewpoint} is to provide a complementary, pedagogical introduction to the effect of stochastic resetting using simple mathematics.

Our starting point is the `inspection paradox,' a famous result in probability theory \cite{SteinDattero,Angus,Ross,MorHarchol,IP1,IP2}. A classic illustration of this paradox was given by William Feller in the second volume of his seminal book on probability theory and applications~\cite{Feller}. Consider the time a passenger waits for a bus. Ideally, buses follow a deterministic process and arrive at fixed time intervals.  In that case a passenger coming to the station at a random time would, on average, wait half the time between consecutive bus arrivals.  However, in reality, time intervals between consecutive bus arrivals fluctuate; some intervals are longer, and some shorter. This creates a sampling bias: The passenger is more likely to arrive during a long time interval than a short one.  Because greater weights are given to longer time intervals, the passenger's average waiting time will always be longer than in the deterministic case. In fact, one can show that a randomly arriving passenger can end up waiting longer, on average, than a passenger who joins the station immediately after a bus departed, which seems paradoxical.

The inspection paradox is not unique to bus arrivals. It is a general phenomenon which emerges from an underlying sampling bias when `measuring' some quantity of interest. For example, the inspection paradox appears in real-life social networks.  It was hence coined the `friendship paradox' by Scott L. Feld \cite{FP}, who observed that most people have fewer friends than their friends have.  For example, imagine picking a social network user at random, and then in a second round, randomly choosing one of his friends. Any user is equally likely to be chosen in the first round, but in the second, users with more friends are over-represented and are thus more likely to be chosen.  This is analogous to the inspection paradox described before, where a user's number of friends is equivalent to the time between consecutive bus arrivals. Context, however, matters: Cohen, Havlin, and  Ben-Avraham showed how the sampling bias of the inspection paradox in social and computer networks can be leveraged to build an efficient immunization strategy which targets highly connected agents~\cite{Ben-Avraham}.  For additional discussion on the inspection paradox and its ubiquity, see the blog \cite{FP1} and lecture \cite{video} by Allen Downey. 

In this \textit{viewpoint}, we will show how stochastic resetting is related to the `inspection paradox,' which in turn will help us formulate `restart criteria' in simple probabilistic terms. To briefly describe the connection, note that processes undergoing renewed dynamics, like the bus arrivals, are also comprised of a series of random time intervals: some shorter, some longer. Similar to the inspection paradox, we are more likely to land in a long interval when choosing a random time to reset the dynamics. Such a resetting mechanism can thus replace long completion times with shorter ones, thereby reducing the average first-passage time of the overall process. Stochastic restart thus works only due to the sampling bias that appears in the inspection paradox. 

For the readers' convenience, we now provide an outline of the paper.  We begin with a detailed discussion of the inspection paradox in Section~\ref{FP}, and show how it is linked to stochastic restart in Section~\ref{InspectionvsPoisson}. We then use this connection to derive the criteria under which restart expedites first passage, for a variety of scenarios such as: (i) simple first passage under restart in Section~\ref{restart}, (ii) first passage of branching processes under restart in Section~\ref{branching}, (iii) first passage under restart with time overheads in Section~\ref{enzyme-cat}, using enzymatic catalysis as an example, and (iv) first passage under restart with space-time coupled returns in Section~\ref{spacetimecoupling}.  We conclude with a summary of the results and additional remarks in Section~\ref{conclusion}.  

In what follows, we will use $f_X(x)$, $\langle X \rangle$, $\sigma_X$, and $CV_X = \sigma_{X} / \langle X \rangle$ to denote, respectively, the probability density function, mean (expectation), standard deviation, and coefficient of variation of a random variable $X$ with non-negative real values.

\section{The inspection paradox}
\label{FP}

The inspection paradox is typically illustrated by the example of a bus stop at which buses regularly arrive. The time $T$ between consecutive bus arrivals is random but has a well-defined average $\langle T \rangle$. 
One might expect that a passenger arriving at the bus stop at some random time will, on average, wait a time $\langle T \rangle/2$ before a bus arrives, i.e. half the average time between two consecutive bus arrivals.  However, this expectation turns out to be incorrect in nearly all cases, except when there are no stochastic fluctuations in arrival times.  Indeed, the average waiting time will always exceed $\langle T \rangle/2$ in the presence of stochastic fluctuations, but the real surprise, or ``paradox,'' emerges from cases where the waiting time exceeds $\langle T \rangle$ itself!  As will be shown shortly, this counter-intuitive result is resolved by a weighted average of waiting times, where greater weights are given to longer times.

\begin{figure}[t]
\centering
\includegraphics[width=\textwidth]{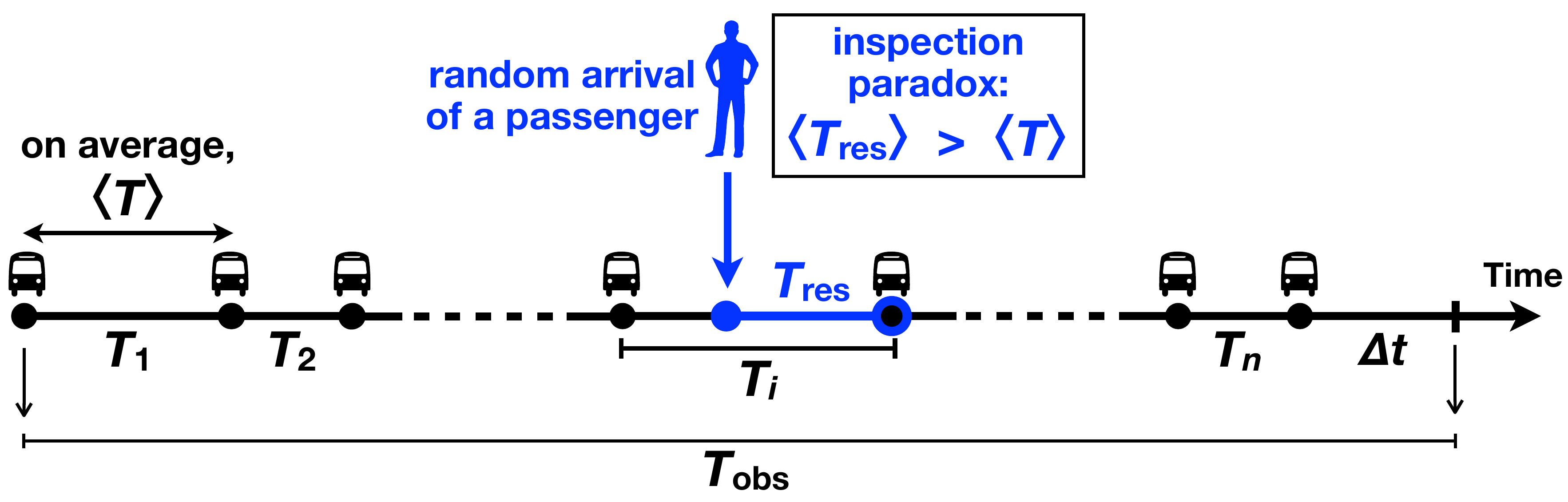}
\caption{An illustration of the inspection paradox.  Consider observing a bus stop for a long time $T_\text{obs}$ and recording the time intervals $T_i$ between the $i$th and $(i+1)$th bus arrivals.  The time intervals $T_i$ vary in length, but on average buses arrive every time interval $\langle T \rangle$.  A passenger arriving at the bus stop at some random time will wait a residual time $T_{\text{res}}$ (highlighted in blue) until the next bus arrival.  On average, how long will the passenger wait?  If stochastic fluctuations in the inter-arrival times are sufficiently large, the average of $T_{\text{res}}$ may exceed $\langle T \rangle$, which can be seen as a manifestation of the inspection paradox.}  
\label{Fig1}
\end{figure}

Figure~\ref{Fig1} illustrates the time intervals $T_i$ between consecutive, i.e. the $i$th and $(i+1)$th, bus arrivals.  We assume that these intervals are independent and identically distributed copies of some generic random variable $T$.  Figure~\ref{Fig1} also illustrates the random arrival of a passenger at the bus stop, who then waits a residual time $T_{\text{res}}$ until the next bus arrives.  Let us first consider the deterministic limit, where the time interval between bus arrivals is fixed and given strictly by $T_i=\langle T \rangle$.  Say that the passenger arrives in the $i$-th interval.  Since the passenger comes at random, s/he is equally likely to arrive at any time within interval $T_i$, i.e. the probability density to arrive at any time point within that interval is uniform.  Therefore, on average, the passenger will arrive at a time $\langle T \rangle/2$ after the $i$th bus, and will wait an average residual time of $\langle T_{\text{res}} \rangle = \langle T \rangle/2$. 

Now let us consider bus arrivals which follow a Poisson process, i.e. assume that the inter-arrival times are exponentially distributed as $f_T(t) = \exp(-T/\langle T \rangle)/\langle T \rangle$.  The exponential distribution is memoryless in the sense that any additional waiting time is independent of the time already spent waiting: $\text{Prob}(T > t+ t' \, | \, T > t) = \text{Prob}(T >t')$. As a result, the residual waiting time $T_{\text{res}}$ of a passenger arriving at a random time following a bus departure comes from the same exponential distribution that governs the generic inter-arrival time $T$, which in turn means that $\langle T_{\text{res}} \rangle=\langle T \rangle$.  Note that in the transition from deterministic to exponentially distributed time intervals, the average residual time doubled from $\langle T \rangle/2$ to $\langle T \rangle$.  This result is already surprising: Even if some time has already passed since the previous bus departed, the passenger must still wait the average inter-arrival time $\langle T \rangle$.

Let us try one more example, with time intervals taken from a Weibull distribution, e.g., with mean $\langle T \rangle$ and density $f_T(t)= \exp (-\sqrt{2 t/\langle T \rangle} ) / \sqrt{2 t \langle T \rangle}$. In this case, as will be shown later, the average residual time $\langle T_{\text{res}} \rangle = (1+\sqrt{5}) \langle T \rangle /2 \approx 1.6 \langle T \rangle $ is in fact \textit{larger} than the average time $\langle T \rangle$ between consecutive bus arrivals! Here the inspection paradox emerges 
:  How is it that, on average, a passenger arriving at a random time must wait considerably longer than the average time between bus arrivals?

The three distributions above highlight three possible scenarios: On average, a passenger will wait a time less than, equal to, or larger than the average bus inter-arrival time.  The latter case is especially surprising -- what distinguishes this case from the other two?  The second moment of the inter-arrival time distribution turns out to be the critical missing parameter.  To see this explicitly, below we reproduce an exact and general expression for the mean residual time~\cite{SteinDattero, Angus, Ross, MorHarchol, IP1, IP2}. An alternate geometrical derivation is provided in~\ref{appendix}.

We denote the bus inter-arrival time distribution as $f_T(t)$, and assume it has a finite mean $\langle T \rangle$ and standard deviation $\sigma_{T}$. Letting $T_{\text{res}}$ stand for the passenger's residual waiting time, we wish to find how its mean $\langle T_{\text{res}}  \rangle$ depends on the properties of $f_T(t)$.  

Say we observe bus arrivals for a long time $T_{\text{obs}}$. Suppose that during the observation time $T_{\text{obs}}$ there are $n+1$ complete bus arrivals occurring at time intervals $T_1,T_2,...,T_i,...,T_n$ as illustrated in Fig.~\ref{Fig1}. Now consider a passenger that arrives at an arbitrary time during $T_{\text{obs}}$. The empirical average of the passenger's residual waiting time can be expressed as a weighted average
\begin{align}
&\overline{T}_{\text{res}}  = \sum_{i=1}^{n} \left(\text{passenger's mean residual time in } T_i \right) \times \left( \text{probability passenger arrives in}~T_i\right) \nonumber \\
& +\left(\text{passenger's mean residual time in}~\Delta t \right) \! \times \! \left( \text{probability passenger arrives in}~\Delta t  \right) \! ,
\label{Eq1}
\end{align}
where $\Delta t $ is the time left between the last bus arrival and the end of the observation period.

By construction, the probability that the passenger arrives at some time $0<t<T_{\text{obs}}$ is uniformly distributed; there is no bias to arrive at one time as opposed to another.  Therefore, the probability that the passenger arrives specifically in time interval $T_i$ is simply the time spanned by that interval, divided by the total observation time:

\begin{equation}
 \text{probability passenger arrives in }T_i = \frac{T_i}{T_{\text{obs}}}  = \frac{T_i}{\left(\sum_{i=1}^{n}T_i \right) +\Delta t}, 
 \label{Eq2}
\end{equation}

Now that we have an expression for the probability that the passenger arrives in interval $T_i$, let us consider the passenger's residual waiting time in interval $T_i$.  Given that the passenger has arrived in interval $T_i$, s/he is \textit{equally likely} to have arrived at any time within this interval. Thus, on average, the residual waiting time is $T_i/2$.  Substituting this result and Eq.~\ref{Eq2} into Eq.~\ref{Eq1}, we obtain
\bea
\overline{T}_{\text{res}} &=& \sum_{j=1}^{n}~\frac{T_j}{2} \cdot \frac{T_j}{\left(\sum_{i=1}^{n}T_i \right) +\Delta t}+\frac{\Delta t}{2}\frac{\Delta t}{T_{\text{obs}}} \nonumber \\
&=& \sum_{j=1}^{n}~\frac{T_j}{2} \cdot \frac{T_j}{\sum_{i=1}^{n}T_i}\cdot \frac{1}{1+\frac{\Delta t}{\sum_{i=1}^{n}T_i}} +\frac{\Delta t}{2 \sum_{i=1}^{n}T_i } \cdot \frac{\Delta t}{1 + \frac{\Delta t}{\sum_{i=1}^{n}T_i}} \, .
\label{Eq3}
\eea
In the limit of long observation times there are infinitely many arrival events ($n \to \infty$), and the term $\Delta t/\sum_{i=1}^{n}T_i$ tends to zero according to the law of large numbers. We thus find 
\bea
\langle  T_{\text{res}}  \rangle &=& \lim_{n \to \infty}~\overline{T} _{\text{res}} =\lim_{n \to \infty}~\frac{1}{2}\frac{\sum_{j=1}^{n} T_j^2}{\sum_{i=1}^{n} T_i}=\lim_{n \to \infty}~\frac{1}{2}\frac{(\sum_{j=1}^{n} T_j^2)/n}{(\sum_{i=1}^{n} T_i)/n}=\frac{\langle T^2 \rangle}{2 \langle T \rangle}.
\label{Eq4}
\eea
The second moment $\langle T^2 \rangle$ can be expressed in terms of the mean and standard deviation $\sigma_T = \sqrt{\langle T^2 \rangle - \langle T \rangle^2}$ of the inter-arrival time distribution:

\bea
\langle  T_{\text{res}}  \rangle = \frac{\langle T^2 \rangle}{2 \langle T \rangle} =  \frac{ \langle T \rangle^2 + \sigma_T^2}{2 \langle T \rangle} = \frac{\langle T \rangle}{2} \cdot \frac{ \langle T \rangle^2 + \sigma_T^2}{\langle T \rangle^2}= \frac{\langle T \rangle}{2}\! \left[ 1+\left(\!\frac{\sigma_T}{\langle T \rangle}\!\right)^{2} \right] \, , \hspace{3mm}
\label{Eq5}
\eea
where the standard deviation-to-mean ratio $\sigma_T/\langle T \rangle$ is simply the coefficient of variation $CV_{T}$ of the inter-arrival time distribution:
\bea
\langle  T_{\text{res}}  \rangle = \frac{\langle T \rangle}{2}\left( 1+CV_{T}^{2} \right).
\label{Eq6}
\eea

The relation above between the mean arrival time $\langle T \rangle$ and the mean residual time $\langle  T_{\text{res}}  \rangle$ results in three possible scenarios:
\begin{itemize}
    \item \emph{Low variability} ($CV_{T} < 1$): For all distributions whose $CV_{T} < 1$, \textbf{the mean residual waiting time is less than the mean inter-arrival time}.  The deterministic case is the extreme limit example: If the inter-arrival time is fixed, $CV_{T}=0$ and therefore $\langle  T_{\text{res}}  \rangle = \langle T \rangle/2$.  Note that the deterministic limit is a lower bound; it is impossible to observe an average residual time less than $\langle T \rangle/2$.
    \item \emph{Marginal case} ($CV_{T} = 1$): By Eq.~\ref{Eq6}, a distribution with $CV_{T} = 1$ will have \textbf{a mean residual  waiting time equal to the mean inter-arrival time}: $\langle  T_{\text{res}}  \rangle=\langle T \rangle$.  One such example arises from the Poisson process, where bus inter-arrival times come from the exponential distribution $f_T(t) = \exp(-t/\langle T \rangle)/\langle T \rangle$. 
    \item \emph{High variability} ($CV_{T} > 1$): If the inter-arrival time distribution has high variability such that $CV_{T} > 1$, then $\langle  T_{\text{res}}  \rangle > \langle T \rangle$.  \textbf{On average, the passenger must wait for a residual time which is longer than the mean inter-arrival time}.  One such example is the Weibull distribution given earlier: $f_T(t)= \exp ( -\sqrt{2 t/\langle T \rangle} ) / \sqrt{2 t \langle T \rangle}$ with mean $\langle T \rangle$. Its coefficient of variation is $\sqrt{5}$, hence the average residual waiting time is $\langle T_{\text{res}} \rangle = (1+\sqrt{5}) \langle T \rangle/2 \approx 1.6 \langle T \rangle $.
\end{itemize}
The high variability case clearly illustrates the inspection paradox. Lurking behind the paradox is the fact that a passenger's random arrival is more likely to happen in a long time interval than in a short one.  This is because the probability of sampling a time interval is proportional to its duration.  In contrast, consider another scenario: The passenger arrives at the bus stop, waits for one bus to arrive and leave, and only then begins the ``residual time stopwatch,'' which runs until the next bus arrival.  In such a case the paradox no longer applies: the distribution of residual times is equivalent to that of inter-arrival times, and no longer suffers from an `inspection bias.'

The inspection bias is sensitive to fluctuations in inter-arrival times, as can be seen from the dependence on the second moment in Eq.~\ref{Eq4}.  The second moment arises from multiplying the probability to arrive in $T_i$ (Eq.~\ref{Eq2}) by the mean residual waiting time there.  Since both factors are proportional to the interval's duration, we obtain $T_i^{2}$, leading to the dependence on the second moment of the inter-arrival time distribution.  In what follows, we will show that the effect of stochastic resetting on the mean completion time of a stochastic process displays a similar sensitivity to fluctuations.  To do so, we first establish a mapping between stochastic resetting and the inspection paradox.  

\section{Stochastic resetting and the inspection paradox}

\begin{figure}[t]
\centering
\includegraphics[width=\textwidth]{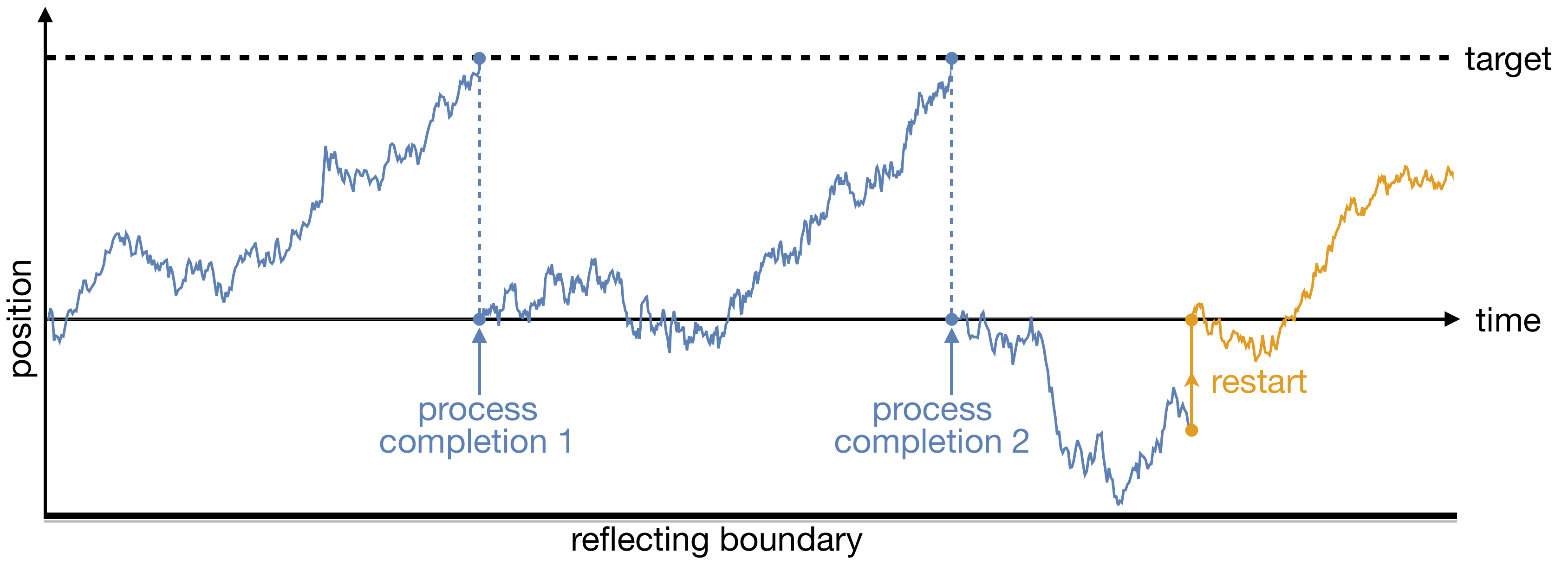}
\caption{Illustration of diffusive search with stochastic resetting. A particle diffuses in a 1D box, searching for a target whose position is indicated by the black dashed line. Upon reaching the target, a new search begins (blue dashed lines). Shown here are two completed search trials, followed by a third which was interrupted by resetting before completion (yellow arrow). Resetting brings the particle back to its starting position, from which it then continues to diffuse (yellow trajectory). Here, resetting is taken to be instantaneous, and non-instantaneous generalizations are treated in Sections~\ref{enzyme-cat} and~\ref{spacetimecoupling}.}
\label{Fig2}
\end{figure}

Similar to bus arrivals in the inspection paradox, one can imagine any process which starts anew upon completion. The classic example in the context of resetting is diffusion. For example, consider a diffusing particle in search of a target, and imagine that the search begins anew every time the particle finds the target. Stochastic resetting interrupts this repeated search process from time to time, placing the particle back at its initial position (see Fig.~\ref{Fig2}). 

A basic question that arises is the following: Can resetting reduce the mean time it takes the diffusing particle to find the target?  Such a question can be asked for any stochastic process which has a start and finish, i.e. can resetting make the process end faster on average? To answer this question, we consider the most widely studied type of stochastic resetting, i.e., one in which resetting occurs at a constant rate $r$. The probability that a resetting event occurs in a short time interval $\Delta t$ is then $r \Delta t$, and resetting events thus follow a Poisson process.  Below we analyze the effect of a small perturbation to the system, that is, a small resetting rate ($r \rightarrow 0$).  In this limit we establish a mapping between stochastic resetting and random inspection (Section~\ref{InspectionvsPoisson}).  We then apply the lessons learned from the inspection paradox, and show that the introduction of resetting reduces the mean completion time of a stochastic process when $CV_{T} > 1$. Using the same mapping, we also derive analogous restart criteria for more complex scenarios which have applications to enzymatic catalysis, animal foraging, and other systems. 

\subsection{Random inspection and low-rate resetting occur with the same uniform probability}
\label{InspectionvsPoisson}

As resetting follows a Poisson process, the number of resetting events $n$ in an observation time $T_{\text{obs}}$ follows a Poisson distribution: $\mathbb{P}(n)=\frac{(r T_{\text{obs}})^n}{n!}~e^{-r T_{\text{obs}}}$. When the resetting rate is low, in the sense that $r T_{\text{obs}}\ll1$, it is most likely that no resetting event occurs during the observation time.  The next most likely scenario is that a single event occurs, and the probability that two or more events occur is negligible.  We will now show that if a single resetting event occurs, it does so with uniform probability in $T_{\text{obs}}$. This is the same probability of a passenger's arrival in the inspection paradox.

Let us consider a time interval of length $T_{\text{obs}}$ which is comprised of three sub-intervals $a_1,b,a_2$ such that $T_{\text{obs}}=a_1+b+a_2$.  The probability that an inspection event, chosen randomly from the entire time interval $T_{\text{obs}}$ with uniform probability, occurs specifically during the interval $b$ is
\begin{align}
\mathbb{P}_{\text{insp}}=\text{Pr}\left( \text{a random inspection occurs in interval $b$} \right)=\frac{b}{T_{\text{obs}}}~.
\end{align}
On the other hand, given that a single resetting event occurs in $T_{\text{obs}}$, the probability $\mathbb{P}_{\text{reset}}$ that it occurs in the interval $b$ is
\begin{align}
\mathbb{P}_{\text{reset}}&=\text{Pr}\left( \text{resetting occurs in interval $b$}~ | ~\text{single resetting event occurs in $T_{\text{obs}}$} \right) \nonumber \\
&=\frac{\text{Pr}\left( \text{`0' events in}~ a_1 ~\text{and `1' event in}~ b~\text{and `0' events in} ~a_2 \right) }{\text{Pr}\left( \text{`1' event in}~ T_{\text{obs}}  \right) }~.
\end{align}
Since the events in the numerator are independent of each other, we can split them as follows:
\bea
\mathbb{P}_{\text{reset}}=\frac{e^{-r a_1}~(r b~e^{-r b})~e^{-r a_2}}{r T_{\text{obs}}~e^{-r T_{\text{obs}}}}=\frac{b}{T_{\text{obs}}}~,
\eea
where we have used the definition of the Poisson distribution above.  Hence $\mathbb{P}_{\text{insp}}=\mathbb{P}_{\text{reset}}$, and we conclude that an inspection event and a resetting event are statistically indistinguishable from each other.  This property will now be used to understand when the introduction of a small resetting rate works to expedite the completion of a stochastic process.

\section{The effect of stochastic resetting}

We now borrow wisdom from the inspection paradox to understand how stochastic resetting affects the mean completion time of a stochastic process.

\subsection{First passage under restart}
\label{restart}

\begin{figure}[t]
\centering
\includegraphics[width=0.9\textwidth]{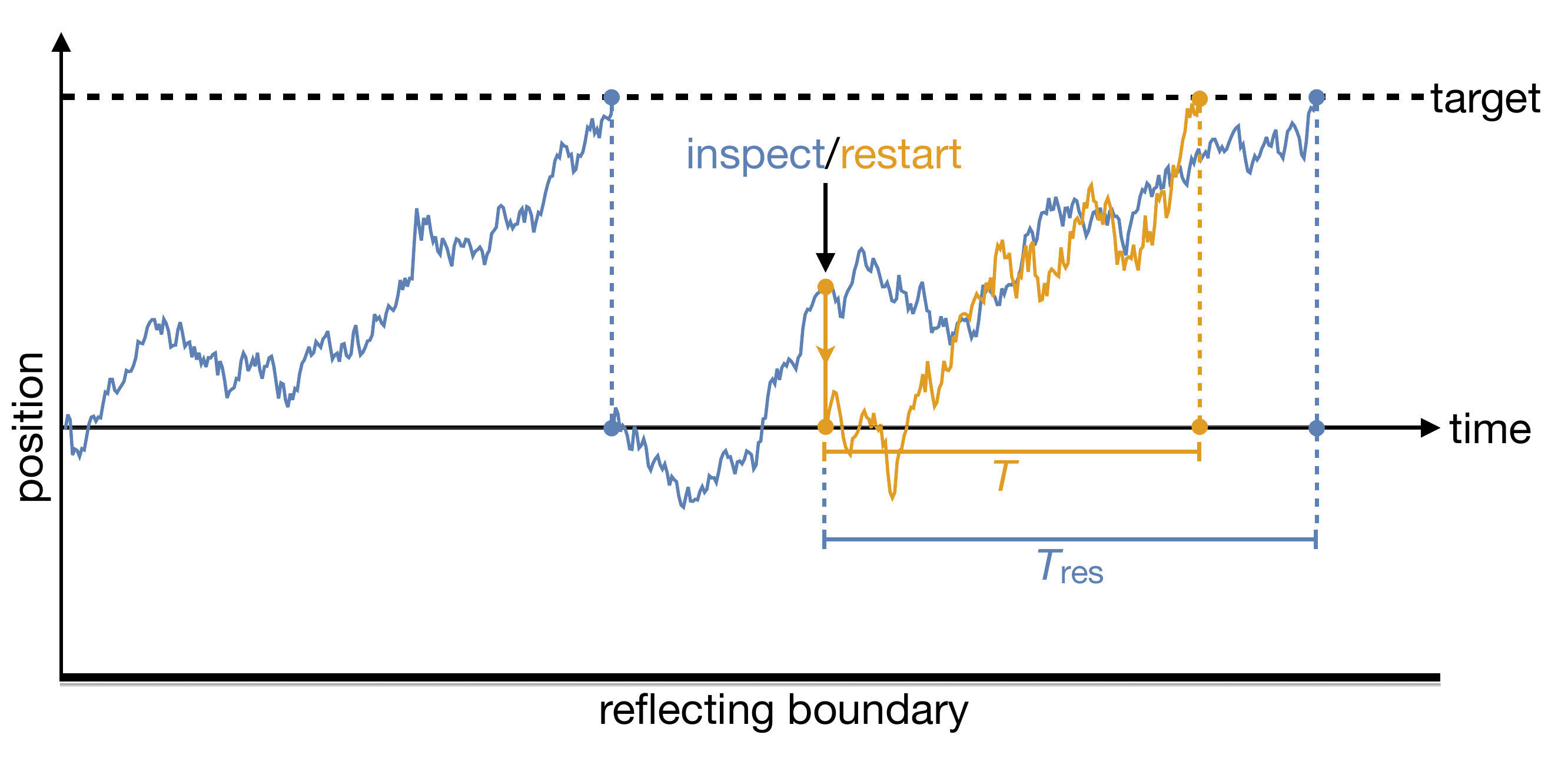}
\caption{Inspection vs. restart.  A typical trajectory subject to inspection (blue), and another trajectory subject to restart (yellow).  The completion time from the inspection point is $T_{\text{res}}$, whereas from the restart point it is $T$. A sufficient condition for resetting to speed up first passage is thus $\langle T \rangle < \langle T_{\text{res}}\rangle$.}
\label{Fig3}
\end{figure}

Since an inspection event and a Poisson restart event occur with the same uniform probability, let us choose a time point randomly in the observation window $[0,T_{\text{obs}}]$ and compare two different trajectories emanating from it: one trajectory which continues unperturbed without restart (as in the inspection paradox), and another which is restarted at that time point, as illustrated in Fig.~\ref{Fig3}.  

Focusing on the first trajectory (without restart), we consider the mean residual time, i.e. the average time remaining for the current search trial to complete.  From the inspection paradox we know that this time is given by $\langle  T_{\text{res}}  \rangle =\langle T^2 \rangle/2 \langle T \rangle$, where $T$ is the stochastic inter-completion time.  Now let us consider the second trajectory which is subject to resetting. Following restart, the average completion time is simply $\langle T \rangle$, since a new search trial began. We now compare the two timescales, and emphasize once more that this can be done since inspection and small-rate resetting occur with the same uniform probability (Section~\ref{InspectionvsPoisson}). This comparison reveals that the introduction of a small resetting rate $r\ll 1/T_\text{obs}$ expedites first passage when
\begin{equation}
    \langle T \rangle < \frac{\langle T^2 \rangle}{2 \langle T \rangle} \, .
    \label{FRUPcriterion1}
\end{equation}
Using Eq.~\ref{Eq5} to rewrite the right-hand side as $\frac{\langle T^2 \rangle}{2 \langle T \rangle} = \frac{\langle T \rangle}{2}\! \left( 1+CV_{T}^{2} \right)$, Eq.~\ref{FRUPcriterion1} simplifies to
\begin{equation}
    CV_{T} > 1,
    \label{nobranching}
\end{equation}
revealing the same result derived in \cite{ReuveniPRL,PalReuveniPRL}. Note that the result obtained is completely general as we did not make use of specific properties of the process described in Fig. \ref{Fig3}. From the perspective of reducing mean completion times, it is thus worthwhile to introduce resetting to a process when its inter-completion time has a coefficient of variation exceeding 1.

The usefulness of this `restart criterion' was demonstrated in many different contexts such as diffusive systems \cite{CV-c1,CV-c2,CV-c2-0,CV-c3,CV-c4,CV-c5,CV-c6,CV-c6-0,CV-c7,CV-c8,CV-c9,CV-c10,comb-1,comb-2,confinement}, active cellular transport \cite{CV-c11}, transport in networks \cite{CV-c12,CV-c12a,CV-c12b}, nonlinear deterministic systems \cite{CV-c13}, intermittent search processes \cite{CV-c14}, and the random acceleration process \cite{CV-c10rap}. Crucially, we turn attention to the broader implications of  Eq.~\ref{nobranching}. For $CV_{T} > 1$ the mean completion time first decreases with the resetting rate as the latter is ramped up from zero (no resetting). The mean completion time either continues in this fashion,  monotonically decreasing as resetting rate increases, or attains a minimum at some critical resetting rate $r^*$. Resetting at this critical rate, which need not be small, can result in significant reduction of the mean completion time, e.g. see  \cite{Restart1,review,expt-1,expt-2,CV-c1,CV-c2,CV-c2-0,CV-c5,CV-c6,CV-c6-0}.

In the analysis above we assumed that the first two moments of the completion time are finite, but similar conclusions apply to situations where the variance and/or mean diverge. Extending the analysis to 
these cases requires more sophisticated tools, but it can nevertheless be shown that the introduction of a small resetting rate always lowers the mean completion time (see \cite{Restart-Biophysics1}).

Finally, we stress that $CV_{T} < 1$ does not necessarily imply that resetting cannot be used to expedite the completion of a stochastic process. While in this case the introduction of a small resetting rate will surely increase the mean completion time, resetting with an intermediate rate may still expedite completion~\cite{Restart-Biophysics2,CV-suff}. Resetting at fixed time intervals (sharp restart) can also expedite certain processes with $CV_{T} < 1$. We refer to \cite{sharp-1,sharp-2,sharp-3} for further discussion.

\subsection{First passage under restart with branching}
\label{branching}

\begin{figure}[t]
\centering
\includegraphics[width=\textwidth]{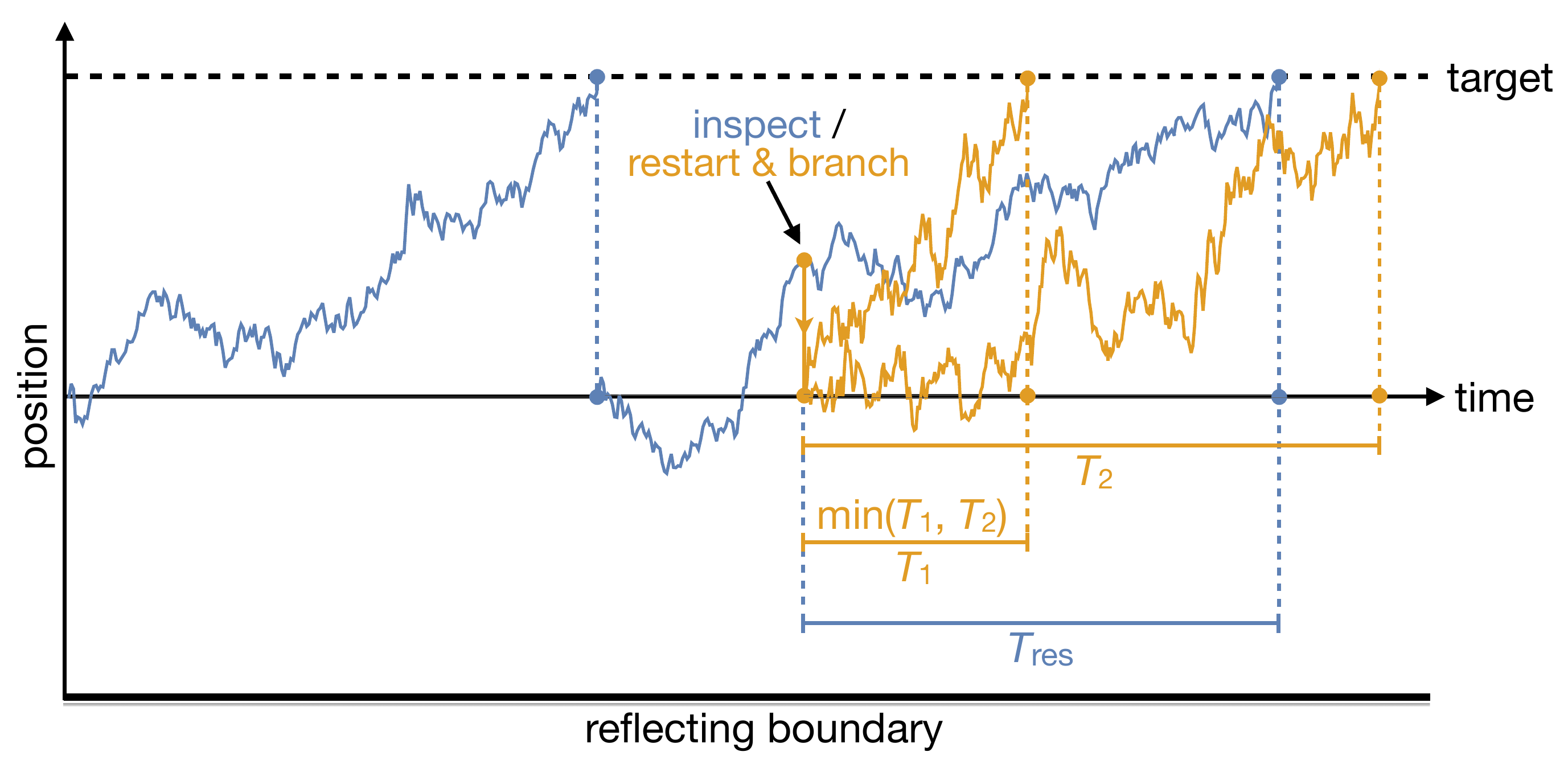}
\caption{Inspection vs. restart with branching. A typical trajectory subject to inspection (blue), and another trajectory subject to restart with branching (yellow). The completion time (time to reach target) from the inspection point is $T_{\text{res}}$, whereas from the `restart \& branch' point it is min$(T_1,T_2)$, i.e., the minimum of two independent and identically distributed copies of a fresh completion time $T$. A sufficient condition for restart with branching to speed up first passage is thus $\langle \text{min}(T_1,T_2) \rangle < \langle T_{\text{res}}\rangle$. 
}
\label{Fig5}
\end{figure}

A different way to break the `$CV=1$ barrier', and expedite completion of processes with $CV<1$, is to augment resetting with an additional 
component: branching. Branching naturally arises in processes such as birth and death \cite{branch-1}, Brownian motion \cite{branch-2}, and epidemic spread \cite{branch-3}. Here we couple it with restart by considering a process which branches into $m$ statistically identical and independent `daughter' processes upon resetting, where $m$ is integer-valued   \cite{PalReuveniBranching,branch-4}. In the same spirit as above, the effect of restart with branching can be understood using insight gained from the inspection paradox (Fig.~\ref{Fig5}).

As before, we ask: In which cases does restart with branching expedite completion of a stochastic process? We again denote the inter-completion time of the underlying process by $T$, and observe that the mean completion time from a random inspection point is the residual time $\langle  T_{\text{res}}  \rangle = \langle T^2 \rangle/2 \langle T \rangle$.  On the other hand, the completion time of the restarted-and-branched process is determined by the fastest branch, i.e. by the minimum of the completion times of $m$ independent and statistically identical branches. Denoting  $T_i$  as an independent and identically distributed copy of the completion time $T$, the completion time of the restarted-and-branched process is $\text{min}(T_1,T_2,...,T_m)$. For restart-and-branching to expedite first passage, we thus require
\begin{equation}
    \langle \text{min}(T_1, T_2, ..., T_m) \rangle < \frac{\langle T^2 \rangle}{2 \langle T \rangle} \, . 
\end{equation}
By Eq.~\ref{Eq5}, the right-hand side can be expressed as $\frac{\langle T \rangle}{2}\! \left( 1+CV_{T}^{2} \right)$, where $CV_{T}$ is the coefficient of variation of $T$, simplifying the criterion to
\begin{equation}
    2 \, \frac{\langle \text{min}(T_1, T_2, ..., T_m) \rangle}{\langle T \rangle} - 1 < CV_{T}^{2} \, . \label{branching_eq2}
\end{equation}
Recognizing the generalized Gini index~\cite{Iddo-Physica} as
\begin{equation}
\mathcal{G}_{m}=1- \frac{\langle \min \left( T_{1},\cdots
,T_{m}\right) \rangle}{\langle T \rangle} \text{ ,}  \label{Gini}
\end{equation}
we arrive at the criterion
\begin{equation}
 CV_{T}^{2} + 2 \mathcal{G}_{m} > 1 \, ,
\end{equation}
which was first derived in~\cite{PalReuveniBranching}.

In the case of no branching ($m=1$), the Gini index vanishes ($\mathcal{G}_{1}=0$) and we recover the criterion $CV_{T} > 1$ (Eq.~\ref{nobranching}). For $m=2$, we obtain $\mathcal{G}_{2}=\mathcal{G}$ which is the conventional Gini index used as a measure of statistical dispersion in economics \cite{Iddo-Physica}.  Higher order indices ($m>2$) generalize the same basic notion.  The above analysis thus reveals how two different measures of statistical dispersion come together to determine whether the introduction of restart with branching expedites the completion of an arbitrary stochastic process~\cite{PalReuveniBranching}. In particular, observe that for $m>1$ and non-deterministic completion times $T$, we have  $\mathcal{G}_{m}>0$. As the coefficient of variation need only obey $CV_{T}^{2} > 1- 2 \mathcal{G}_{m}$, we conclude that restart with branching can expedite completion of processes that simple restart cannot.

\subsection{Generalization to restart with time overheads: Enzymatic catalysis}
\label{enzyme-cat}

\begin{figure}[h]
\centering
\includegraphics[width=0.95\textwidth]{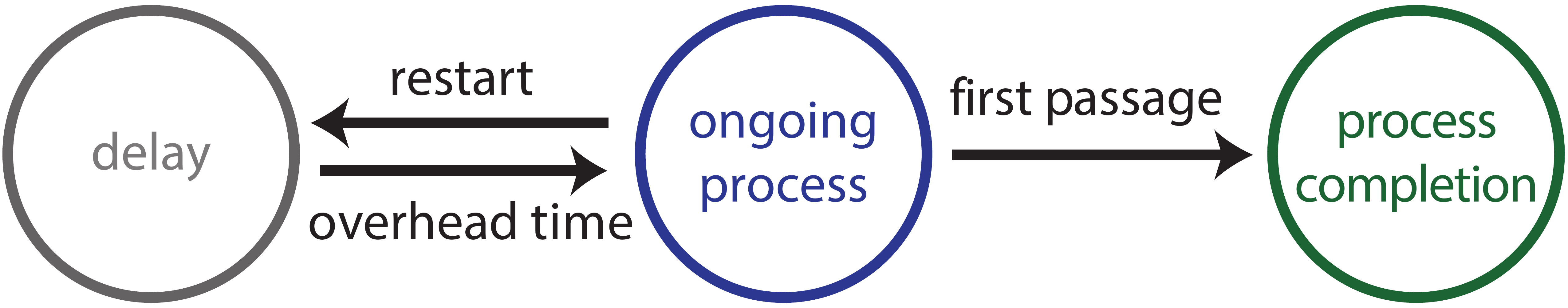}
\caption{A schematic illustration of first passage under restart with time overheads (delays).  Unlike the previously considered case of instantaneous restart, here a time delay is incurred upon resetting, which is typical in realistic scenarios.}
\label{time-overheads}
\end{figure}

\noindent
Thus far we have assumed that resetting takes place instantaneously. However, in reality one often needs to wait some time (random or deterministic) after a process is stopped and before it starts over~\cite{ReuveniPRL,Restart-Biophysics1,Restart-Biophysics2,Restart-Biophysics3,Restart-Biophysics4,Delay1,Delay2,Delay3}, as illustrated in Fig.~\ref{time-overheads}. For example, imagine restarting a computer software or algorithm. Before the program is run again, some time is spent on, e.g., initializing certain variables and loading the program onto available memory. Similar delays 
also occur in natural systems, such as in enzymatic catalysis, which is discussed below.

\begin{figure}[h]
\centering
\includegraphics[width=0.9\textwidth]{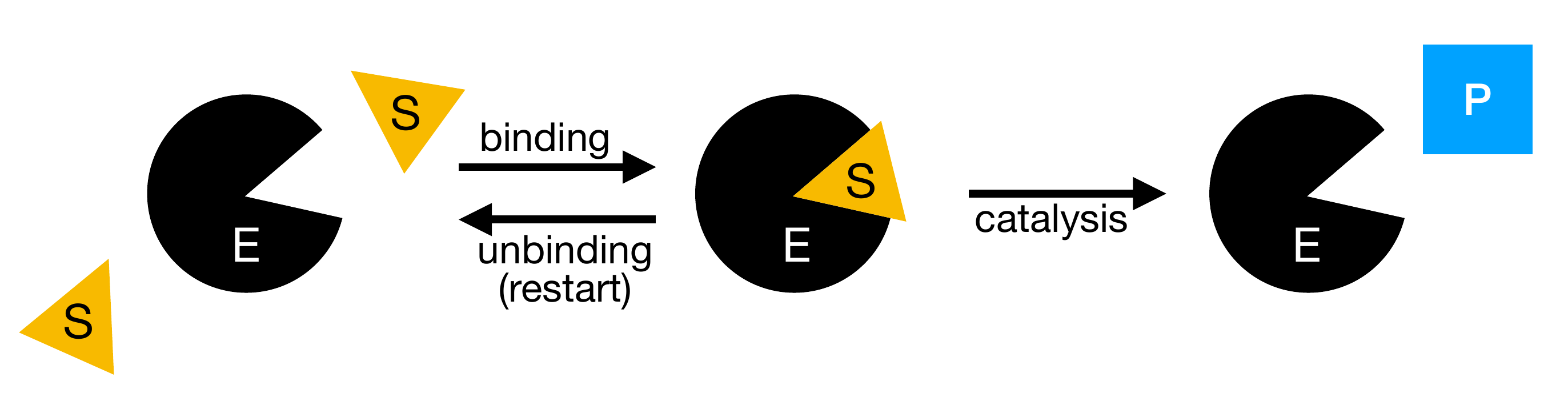}
\caption{A schematic illustration of the Michaelis-Menten model of enzymatic catalysis. Here, $E$ denotes the enzyme (black), $S$ the substrate (light orange), and $P$ the product (blue): $E + S \rightleftharpoons ES \rightarrow E + P$.  The unbinding of the substrate from the enzyme before catalysis takes place can be considered a `restart' event (bottom left arrow).}
\label{enzyme-cat-1}
\end{figure}

Enzyme-catalyzed reactions are essential to life on earth~\cite{Biochemistry,Lehninger}. They are often described by the Michaelis-Menten model~\cite{Michaelis-Menten}, which is a  cornerstone of chemical kinetics. In this model, an enzyme $E$ reversibly binds a substrate $S$ to form a metastable complex $ES$. The substrate can then be converted by the enzyme to form a product $P$ via catalysis or, alternatively, unbind as illustrated in Fig.~\ref{enzyme-cat-1}. It is apparent that the schemes in  Figs.~\ref{time-overheads} and~\ref{enzyme-cat-1} are identical aside from labels: Catalysis can be viewed as a first-passage event, unbinding as restart, and binding is always accompanied by a time delay. Thus, the Michaelis-Menten model of enzymatic catalysis describes first-passage under restart with time overheads~\cite{Restart-Biophysics1}.

\newpage 

Since chemical reactions at the single molecule level are inherently stochastic, the binding, unbinding and catalysis processes illustrated in Fig.~\ref{enzyme-cat-1} are all random. We thus describe the binding and catalysis times by some generic random variables which we denote as $T_\text{on}$ and $T$, respectively. Similar to before, we ask: How does unbinding affect the mean time it takes for an enzyme to catalyze the conversion of substrate molecule to a product molecule?  Equivalently, how does resetting followed by a time delay, affect the mean completion time of a generic stochastic process?

\begin{figure}[t]
\centering
\includegraphics[width=\textwidth]{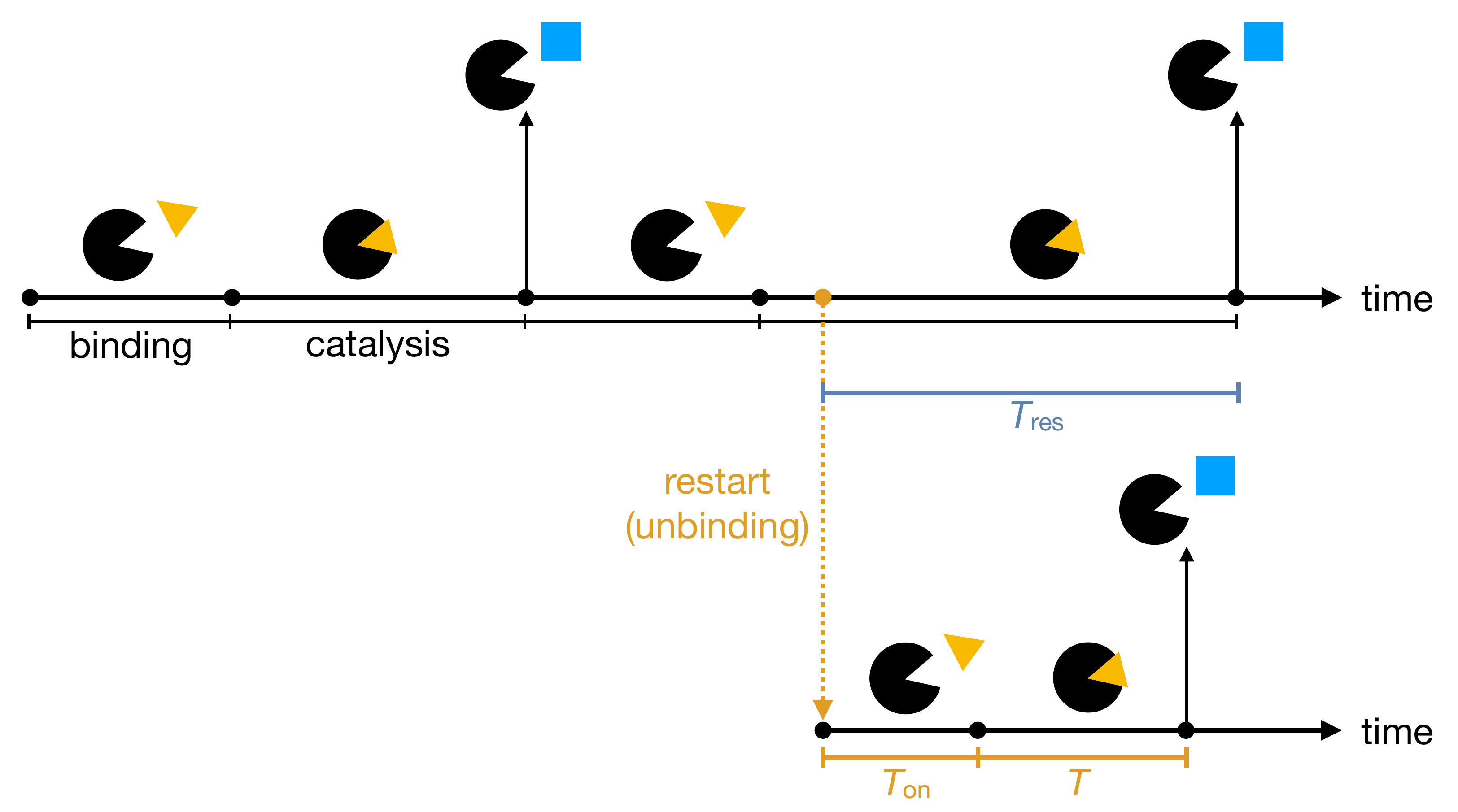}
\caption{Resetting with time overheads naturally occurs in enzymatic catalysis. Unbinding `restarts' the catalysis process whose random duration is otherwise $T$. This results in a delay of random duration $T_{\text{on}}$, the time it takes the enzyme to bind to another substrate.  The introduction of a small unbinding (restart) rate will expedite reaction completion when $\langle T_{\text{res}} \rangle > \langle T_{\text{on}} \rangle + \langle T \rangle$, i.e., when the mean residual time for reaction completion is larger than the average time required to complete a fresh reaction cycle.}
\label{enzyme-cat-2}
\end{figure}

To answer this question, first note that  unbinding can only occur during the catalysis phase, and consider once again the effect of a small unbinding (resetting) rate. Recalling the mapping with the inspection paradox, we compare two scenarios (Fig.~\ref{enzyme-cat-2}): In the first, an observer inspects the system at a random time, but only during catalysis. As we already know, this random inspection will reveal an average residual time of $\langle  T_{\text{res}}  \rangle = \langle T^2 \rangle/2 \langle T \rangle$ till the reaction completes.  In the second scenario, unbinding occurs which immediately releases the enzyme $E$ from the substrate $S$ to start a new reaction cycle (binding followed by catalysis). Since the unbinding rate is very small by construction, the probability that two unbinding events occur during the observation time is negligible.  Hence, on average, following an unbinding event it will take $\langle T_{\text{on}} \rangle+\langle T \rangle$ for the reaction to complete. 

Comparing the two scenarios above, we see that 
\bea
\langle T_{\text{res}} \rangle >  \langle T_{\text{on}} \rangle+\langle T \rangle,
\eea
ensures that the introduction of a small unbinding rate will lower the mean turnover time of the reaction. Rearranging this relation, we arrive at 
\bea
CV_{T}^{2} > 1+\frac{2 \, \langle  T_\text{on} \rangle}{\langle  T \rangle}.
\label{pathindependent}
\eea
Note that when the average binding time is negligible (no delays),  $\langle T_{\text{on}} \rangle$ vanishes and we recover the criterion $CV_{T} > 1$ of Eq.~\ref{nobranching}. Otherwise, the coefficient of variation must be larger than unity, as finite binding times (delays) result in a penalty whose magnitude is twice the ratio of the average binding time to the average completion time of the catalytic process. We nevertheless see that when relative stochastic fluctuations are large enough, restart can expedite completion even in the presence of such time overheads.

\subsection{Generalization to restart with space-time coupled overheads}
\label{spacetimecoupling}
\begin{figure}[t]
\centering
\includegraphics[width=\textwidth]{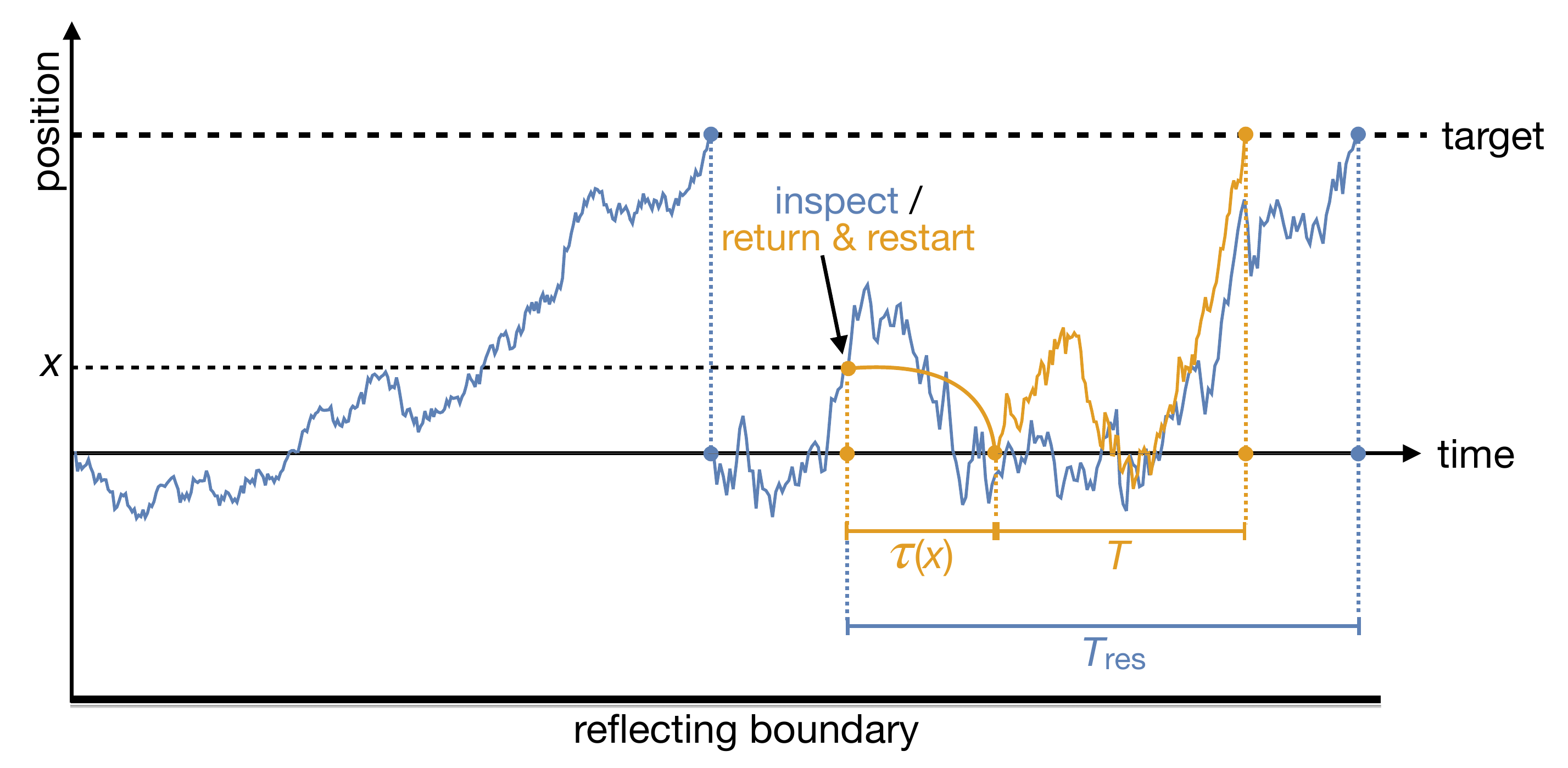}
\caption{In realistic descriptions of stochastic motion with resetting, time overheads are often coupled to the spatial location. This is illustrated above by the first-passage trajectory of a particle subject to resetting with non-instantaneous returns (light orange), in contrast to that of a diffusing particle without resetting (blue). In the former case, return to the origin takes some time $\tau(x)$ which depends on the particle's position $x$ at the resetting moment via, e.g., the return path and speed. The introduction of a small resetting rate expedites first-passage when $\langle  T_{\text{res}}  \rangle > \langle \tau(x)\rangle+\langle T \rangle $, i.e. when the mean residual time to reach the target is larger than the average time 
to return to the origin and diffuse to the target from there.}
\label{home-search}
\end{figure}

\noindent
In the previous subsection, we discussed first passage under restart with generic time overheads.  It was implicitly assumed that delays which follow resetting events are completely decoupled from the state of the process at the moment of resetting.  This assumption can be justified for, e.g., enzymatic catalysis, where delays due to binding times are mainly affected by the enzyme-substrate collision rate rather than the internal state of the enzyme at the moment of unbinding (reset).  In contrast, consider again diffusion with resetting: While we previously assumed resetting to be instantaneous, in reality returning a particle to the origin takes time~\cite{expt-1}.  Furthermore, returning a particle from afar will take longer than returning one from nearby. Any realistic description of stochastic motion with resetting thus entails a time delay whose duration depends on the particle's location at the moment of resetting (Fig.~\ref{home-search}). To that end, a general approach to resetting with non-instantaneous, space-time coupled returns was introduced in~\cite{non-inst-det-2,non-inst-det-3,HRS}.

We now derive a criterion for when restart expedites first-passage in the presence of space-time coupled delays. As before, we denote the completion times of the underlying stochastic process by the random variable $T$.  Resetting the process incurs a delay time $\tau(\vec{x})$ to return the particle to its initial position, after which the process restarts.  Note that $\tau(\vec{x})$ is a general deterministic function of the particle's position $\vec{x}$, which itself is a random variable.  Upon return, the process renews.  This search-reset-return cycle repeats itself until the process is completed, e.g. a target is reached, at some point during the search stage.  

For an unperturbed process inspected at a random time point, we know that the mean residual time for completion is given by $\langle  T_{\text{res}}  \rangle = \langle T^2 \rangle/2 \langle T  \rangle$ (Fig.~\ref{home-search}). On the other hand, resetting at a random time point returns the particle to its initial location over a course of time $\tau(\vec{x})$, and the process starts over. Since we consider a small resetting rate, no more than a single resetting event occurs during the observation time. Thus, on average, the time to complete the process following a resetting event is $\langle \tau(\vec{x})\rangle+\langle T \rangle$, where the averaging on $\langle \tau(\vec{x}) \rangle$ is  over the particle's random position $\vec{x}$ at the random time of resetting.

Comparing the timescales above, we see that the introduction of a small resetting rate expedites completion when 
\bea
\langle \tau(\vec{x})\rangle+\langle T \rangle< \langle  T_{\text{res}}  \rangle. 
\label{pathdependent0}
\eea
The mean return time on the left-hand side can be written as \cite{HRS}
\bea
\langle \tau(\vec{x}) \rangle = \int_{\mathcal{D}}~d\vec{x}~\tau(\vec{x}) \phi(\vec{x}) = \frac{1}{\langle T  \rangle} \int_{\mathcal{D}}~d\vec{x}~\tau(\vec{x}) \int_0^\infty ~dt~ G(\vec{x},t) ,
\eea
where $\mathcal{D}$ denotes the spatial domain of interest for the process. The averaging of $\tau(\vec{x})$ is done with respect to the probability measure $\phi(\vec{x})=\frac{1}{\langle T  \rangle}  \int_0^\infty ~dt~ G(\vec{x},t)$ defined in terms of the time-dependent propagator $G(\vec{x},t)$, which gives the probability density to find the particle at position $\vec{x}$ at time $t$ in the absence of resetting. Note that the propagator is not normalized to unity due to absorbing boundaries or targets at which the process terminates; its integral over the domain $\mathcal{D}$ yields the survival probability $Pr(T>t)$. The normalization condition for the measure $\phi(\vec{x})$ then  reads $\int_{\mathcal{D}}~d\vec{x}~\phi(\vec{x})=\frac{1}{\langle T \rangle}\int_0^\infty ~dt~\int_{\mathcal{D}}~d\vec{x}~G(\vec{x},t)=\frac{1}{\langle T \rangle} \int_0^\infty ~dt~ Pr(T>t)=1$.

Rearranging Eq.~\ref{pathdependent0}, we arrive at 
\bea
CV_{T}^{2} > 1+\frac{2 \, \langle \tau(\vec{x}) \rangle}{\langle T  \rangle},
\label{pathdependent}
\eea
which for negligible return times again boils down to Eq.~\ref{nobranching}. The condition in Eq.~\ref{pathdependent} reveals that search with returns to the initial location can outperform free search in conditions of high uncertainty. This is evident from the left-hand side of Eq.~\eref{pathdependent} which quantifies the relative magnitude of fluctuations, or uncertainty, around the mean first-passage time of the underlying search process. Indeed, when these fluctuations are large the introduction of resetting will result in a shorter time to the target, despite delays incurred by space-time coupled returns \cite{HRS}. 

\section{Summary and concluding remarks}
\label{conclusion}
In this \textit{viewpoint}, we have put forward the connection between the inspection paradox from probability theory and stochastic resetting, a recent focal point in non-equilibrium statistical physics, stochastic processes, chemistry and other interdisciplinary fields. A key observation is that restart can often expedite the completion time of a stochastic process. This important property, however, is not always true, and understanding when it holds has been a central goal in the field.  In this perspective, we provided simple physical interpretations of such restart criteria by borrowing wisdom from the famous inspection paradox of probability theory.  We showed how results from the inspection paradox can be mapped onto first-passage under instantaneous restart and extended to more complex scenarios. 

The inspection paradox highlights the inherent sampling bias that accompanies large fluctuations, and we can use that sampling bias to our advantage with restart.  The greater the fluctuations in the stochastic process, e.g in inter-arrival times of buses, the more likely we are to land in a long interval and the longer we will wait for process completion.  The magnitude of the fluctuations need not be exceedingly large to see such sampling bias: As soon as the coefficient of variation exceeds 1 -- which marks the marginal case of a Poisson process ($CV^{\text{Poiss}}_T = 1$) -- the sampling bias becomes significant and pushes the average waiting time to exceed the average interval duration $\langle T \rangle$.  Restart provides a mechanism to extract us from long time intervals and mitigate the sampling bias, which effectively reduces our waiting time.  The effect of restart becomes more pronounced the greater the fluctuations, and can still reduce waiting times even in the presence of time overheads incurred by resetting.

Two central results for resetting with time overheads are Eqs.~\ref{pathindependent} and \ref{pathdependent}.  Their respective time delay contributions can be added to yield a single criterion under which Poissonian restart expedites first passage:
\begin{equation}
    CV_{T}^{2} > 1 + \underbrace{\frac{2 \, \langle T_{\text{on}} \rangle}{\langle T \rangle}}_{\text{regular delays}} + \underbrace{\frac{2 \, \langle \tau(\vec{x}) \rangle}{\langle T \rangle}}_{\text{\parbox{2cm}{\centering space-time \\[-4pt] coupled delays}}}.
    \label{unifiedcriterion}
\end{equation}
Recall that $CV_T$ is the coefficient of variation of the underlying process (described by time intervals of random duration $T$), and that $T_{\text{on}}$ is a random delay time which is independent of space. In addition, $\tau(\vec{x})$ is a time delay which is a deterministic function of the random resetting position $\vec{x}$. This time accounts for the space-time coupled returns.

When restart is instantaneous, as considered in Section~\ref{restart}, then $T_{\text{on}} = \tau(\vec{x}) = 0$ and we recover the original, simpler result of $CV_{T} > 1$.  Meanwhile, for space-independent time delays, $\tau(\vec{x}) = 0$ and so we recover the previously obtained criterion of Eq.~\ref{pathindependent}.  Similarly, for space-time coupled delays, $T_{\text{on}} = 0$ which yields Eq.~\ref{pathdependent}.  In the case that both space-independent and space-dependent time delays are present, we can simply add their contributions as displayed in Eq.~\ref{unifiedcriterion}.  This unified criterion can thus serve as a guide for those who wish to determine whether resetting will expedite process completion in their specific system. \\

\textbf{Acknowledgments}. The authors thank Yuval Scher, Maxence Arutkin, Ofek Lauber, and Yair Shokef for commenting on early versions of this manuscript. A.P. gratefully acknowledges support from the Raymond and Beverly Sackler postdoctoral scholarship and the Ratner Center for Single Molecule Science at Tel Aviv University. S.K. was supported by the Zuckerman STEM postdoctoral fellowship. S. R. acknowledges support from the Azrieli Foundation, from the Raymond and Beverly Sackler Center for Computational Molecular and Materials Science at Tel Aviv University, and from the Israel Science Foundation (grant No. 394/19).\

\appendix
\section{Alternate proof of Equation 4}

\label{appendix}

\begin{figure}[h]
\centering
\includegraphics[scale=0.6]{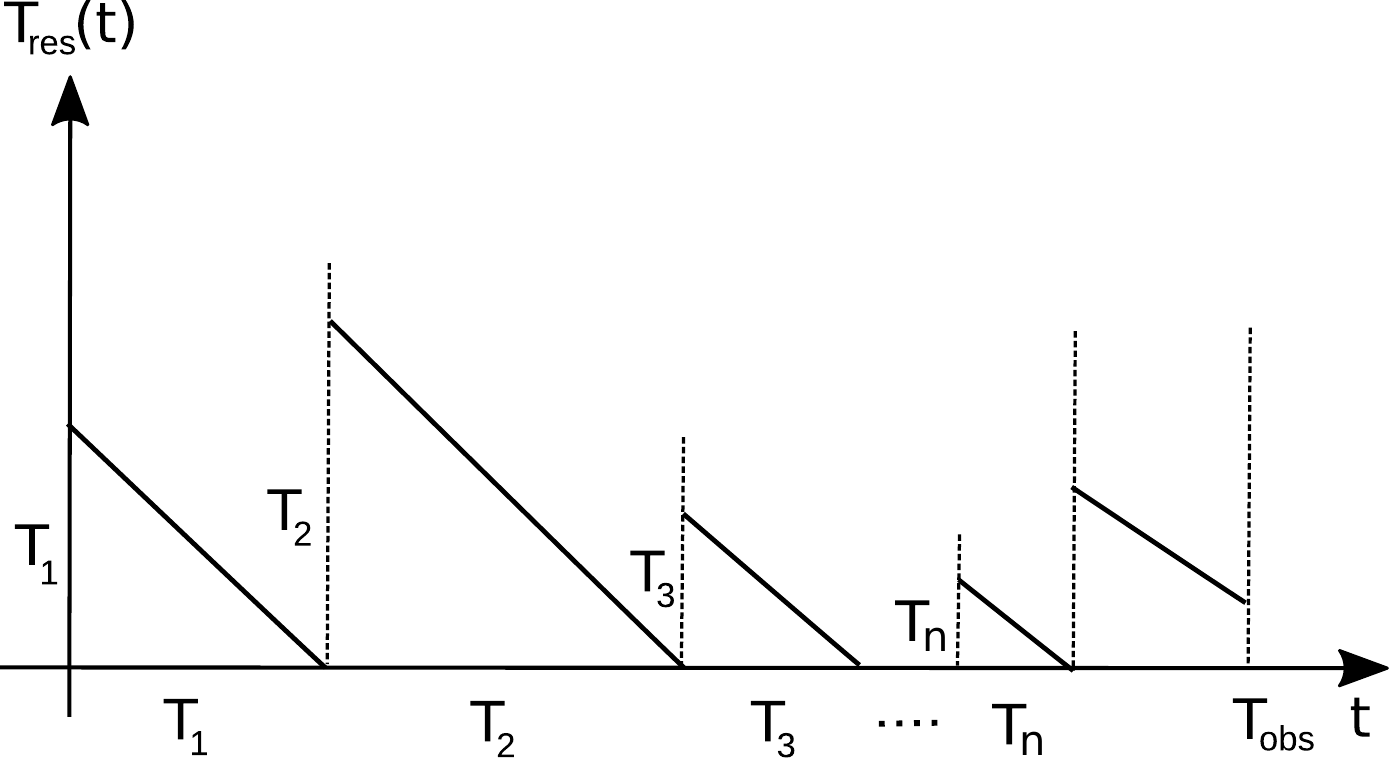}
\caption{A stochastic trajectory of the residual waiting time $T_{\text{res}}(t)$ as a function of passenger's arrival time $t$. The process is observed till a fixed time $T_\text{obs}$.}
\label{Figgeo}
\end{figure}

Consider a renewal process as depicted in Fig.~\ref{Figgeo} which consists of a sequence of time intervals $T_1,T_2,...,T_i,...,T_n$ between bus arrivals, and further assume that these times are independent and identically distributed copies of a generic random variable $T$ whose mean and variance are finite. Recall that $T_{\text{res}}$ is the residual time that a passenger arriving at random must wait until the next bus arrival, i.e. the time which remains in the interval $T_i$ in which s/he arrived. Letting $t$ denote the arrival time of the passenger, we see $T_{\text{res}}(t)$ is bounded by $T_i$ from above, creating a `sawtooth' behavior of $T_{\text{res}}$ as shown in Fig.~\ref{Figgeo}. The empirical average of the residual waiting time is given by
\bea
\overline{T}_{\text{res}} =
\frac{\int_0^{T_{\text{obs}}} ~T_{\text{res}}(t)~dt}{T_{\text{obs}}},
\eea
which is simply the area under the curve spanned by $T_{\text{res}}(t)$ in Fig.~\ref{Figgeo}, normalized by the observation time.  We can expand the numerator as
\bea
\overline{T}_{\text{res}} =
\frac{\sum_{i=1}^n~\int_0^{T_i}dt~ T_{\text{res}}(t)+\int_{\sum_{i=1}^n T_i}^{T_\text{obs}}dt~ T_{\text{res}}(t)}{T_{\text{obs}}},
\eea
where the $i$-th integral  $\int_0^{T_i}~dt~T_{\text{res}}(t)$ gives the area of the i-th triangle in Fig.~\ref{Figgeo} ($1 \leq i \leq n$). Having a height and base of length $T_i$, the area of each such triangle is $T_i^2/2$, and we denote the area of the last integral (which is not necessarily a triangle) as $\Delta$. Thus, we obtain
\bea
\overline{T}_{\text{res}} =
\frac{\frac{1}{2}\sum_{i=1}^n~T_i^2+
\Delta}{\sum_{i=1}^n~T_i+
\Delta t}.
\eea
Now, let us divide both the numerator and denominator by $n$.  In the limit of long observation times and infinitely many arrival events ($n \to \infty$), the term $\Delta t/n$ tends to zero, and the area contribution from the last interval $\Delta/n$ can also be neglected. Thus, we obtain the mean residual time 
\bea
\langle  T_{\text{res}}  \rangle &=& \lim_{n \to \infty}~\frac{\frac{1}{2n}\sum_{i=1}^n~T_i^2+
\Delta/n}{\frac{1}{n}\sum_{i=1}^n~T_i+
\Delta t/n} = \frac{\langle T^2 \rangle/2}{\langle T \rangle},
\eea
which concludes the proof.\\

\newpage
\noindent \textbf{References}\\

{}

\end{document}